# Lattice Opening Upon Bulk Reductive Covalent Functionalization of Black Phosphorus

Stefan Wild, Michael Fickert, Aleksandra Mitrovic, Vicent Lloret, Christian Neiss, José Alejandro Vidal- Moya, Miguel Ángel Rivero-Crespo, Antonio Leyva-Pérez, Katharina Werbach, Herwig Peterlik, Mathias Grabau, Haiko Wittkämper, Christian Papp, Hans-Peter Steinrück, Thomas Pichler, Andreas Görling, Frank Hauke, Gonzalo Abellán* and Andreas Hirsch*

[a] S. Wild, M. Fickert, Dr. A. Mitrovic, V. Lloret, Dr. F. Hauke, Dr. G. Abellán, Prof. A. Hirsch
[b] Chair of Organic Chemistry II and Joint Institute of Advanced Materials and Processes (ZMP); Friedrich-Alexander-Universität Erlangen-Nürnberg (FAU); Nikolaus-Fiebiger Straße 10, 91058 Erlangen and Dr.-Mack Straße 81, 90762 Fürth (Germany)
E-mail: andreas.hirsch@fau.de and gonzalo.abellan@fau.de
[c] Dr. C. Neiss, Prof. A. Görling
Lehrstuhl für Theoretische Chemie and Interdisciplinary Center of Molecular Materials (ICMM); Friedrich-Alexander-Universität Erlangen-Nürnberg (FAU); Egerlandstraße 3, 91058 Erlangen (Germany)
[d] Dr. J. A. Vidal-Moya, M. A. Rivero-Crespo, Dr. A. Leyva-Pérez
Instituto de Tecnología Química. Universidad Politècnica de València–Consejo Superior de Investigaciones Científicas. Avda. de los Naranjos s/n, 46022, Valencia (Spain)
[e] K. Werbach, Prof. H. Peterlik, Prof. T. Pichler
Faculty of Physics, University of Vienna; Strudlhofgasse 4, 1090 Vienna (Austria)
[f] M. Grabau, H. Wittkämper, Dr. C. Papp, Prof. Dr. H.-P. Steinrück
Lehrstuhl für Physikalische Chemie II, FAU, Egerlandstraße 3, 91058 Erlangen (Germany)
[g] Dr. G. Abellán
Instituto de Ciencia Molecular (ICMol), Universidad de Valencia, Catedrático José Beltrán 2, 46980, Paterna, Valencia (Spain)

**Abstract**

The chemical bulk reductive covalent functionalization of thin layer black phosphorus (BP) using BP intercalation compounds has been developed. Through effective reductive activation, covalent functionalization of the charged BP is achieved by organic alkyl halides. Functionalization was extensively demonstrated by means of several spectroscopic techniques and DFT calculations, showing higher functionalization degrees than the neutral routes.

Since 2014, two-dimensional (2D) Black Phosphorus (BP) has attracted enormous attention throughout the scientific community due to its high p-type charge carrier mobility and its tunable direct bandgap.[1–9] In contrast to graphene, BP exhibits a marked puckering of the $sp^3$ structure, constituting a two-dimensional σ-only system, involving one lone electron pair at each P atom. Whereas its outstanding physical and materials properties have been intensively investigated, its chemistry remains almost unexplored.[10–12] Indeed, a first-series of non-covalent functionalization protocols has been reported, mainly focused on improving the intrinsic instability of BP against water and oxygen.[13–17] Beyond these approaches, the covalent functionalization of the interface is one of the most promising routes for the fine-tuning of the chemical and physical properties of 2D nanomaterials.[18,19] In this sense, only a few recent works on single-flake chemistry with diazonium salts[20], or wet-chemistry on previously exfoliated flakes with nucleophiles[21–23] or carbon-free radicals[24] have been reported so far. This is probably due to the intrinsic low degree of reactivity towards these reactions of neutral BP and the difficulties associated to overcome the huge van der Waals energy stored within a BP crystal, thus blocking the direct functionalization of BP. Along this front, a bulk wet-chemical derivatization sequence remains to be found. Moreover, an unambiguous determination of the covalent binding and its influence in the chemical structure of the P$_4$ units is required to systematically explore the characteristics of BP reactivity.

For facing these challenges we took advantage of the well- known reductive graphene chemistry using graphite intercalation compounds (GICs).[18,25–27] As a first success in this direction, we have recently reported the preparation of BP intercalation compounds (BPICs) with alkali metals (K and Na).[28] This paves the way for the exploration of the reductive route using activated negatively charged BP-*ite* nanosheets and electrophiles (*E*) as covalent reaction partners.

Herein, we provide the first real proof for covalent binding in BP with alkyl halides using a battery of characterization techniques. Furthermore, density functional theory (DFT) calculations were carried out to rationalize our results, providing a deep understanding of the covalent derivatization of BP. This thorough study reveals for the first time the lattice opening in BP, absent in graphene, which is a P-characteristic phenomenon in reduction chemistry, with the extreme case of K$_3$P formation with all P-P bonds cleaved.[28] First, we investigated the *in-situ* treatment of a potassium-BPIC upon the addition of an electrophilic functionalization reagent (Scheme 1) followed by Raman spectroscopy.[28] Therefore, we slowly evaporated hexyliodide onto the previously synthesized BPIC KP$_6$, under ultra-high vacuum (UHV) conditions and measured constantly Raman spectra in order to monitor the course of the functionalization process *in situ*. The BP was intercalated in a glovebox under Argon atmosphere and transferred into the UHV-reaction chamber (SI 1). As depicted in Figure 1a), highlighted in grey, a series of new distinct Raman modes arise below 400 cm$^{-1}$ concomitantly with the increasing amounts of the evaporated electrophilic hexyliodide. More specifically, a sharp peak at 145 cm$^{-1}$, a broad shoulder around 210 cm$^{-1}$, two broad features between 270 and 320 cm$^{-1}$ and a small peak between the A$_g^1$ and B$_{2g}$ peak at 405 cm$^{-1}$ can be clearly observed, strongly suggesting the modification of the 2D- BP lattice with the formation of a P–C bond. We carried out DFT calculations simulating the covalent attachment of a methyl group to BP and calculating the expected Raman spectra.

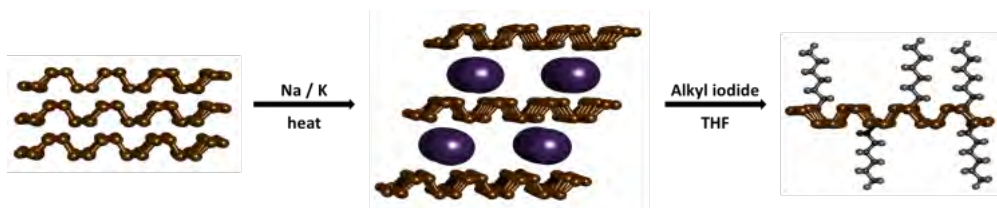

**Scheme 1.** General reaction course showing the reductive covalent functionalization of BP. Pristine BP is intercalated with an alkali metal in solid state under controlled heating and afterwards the activated BPIC is dispersed in THF and reacted with an electrophilic trapping reagent.

We used a 4 x 3 super cell of BP, with one methyl group added. We considered several binding scenarios, including the formation of positively charged phosphonium salts, the influence of K and I, the functionalization degree and the monotopic/antaratopic functionalization. Interestingly, the resulting structures, after a full geometry relaxation was allowed, repeatedly exhibit one localized unpaired electron as well as an elongated P–P bond next to the P–C bond, concretely from 2,26 to 2,83 Å, leading to a lattice opening (Scheme 2, SI 2). As this resulting radical species is expected to be highly reactive, we also considered saturation by iodine or potassium, which would remove spin polarization (radical character) of the methylated BP single layer (SI 3). Figure 1b shows the calculated Raman spectra for a methylated BP single layer in the presence of potassium (further calculations can be found in SI 4–6), which takes into account that the vapor-solid reaction was not quenched in the *in situ* experiment, and therefore the potassium definitely remains in the resulting compound. According to our calculations the intensities –especially of the low-frequency bands– are quite sensitive to the degree of functionalization, charging and presence of counter ions (see Figs. SI 10, SI 4-6). In this sense, the predicted Raman spectrum of a methylated BP monolayer saturated with potassium is in good agreement with our experimental results (to correct for systematically too low frequencies the calculated spectra were shifted to match the experimental $A_g^2$ band at 467 cm$^{-1}$). Moreover, we obtain the same results when using Na- instead of K-BPICs (Figure SI 7). This is in clear contrast with GICs chemistry, in which the Na- intercalation compounds cannot be achieved without using Na$^+$(G$_x$)$_y$-complexes, where G accounts for linear ethylene glycol dimethyl ether homologues ("glymes").[29]

In the next step, we focused on the bulk reductive functionalization of BP using the wet chemical approach.[18,30] For this, pristine BP was intercalated with sodium or potassium to yield the respective BPIC (NaP$_6$ or KP$_6$), which afterwards was dispersed applying ultra-sonication in purified THF, and finally quenched with alkyl halides (Scheme 1). After filtration in the glovebox, the reaction product was obtained in form of a dark grey powder. Statistical Raman spectroscopy (SRS) using an excitation wavelength of 633 nm revealed the appearance of a new band at around 145 cm$^{-1}$, which was also observed in the *in-situ* experiment, and new Raman modes in the 250–300 cm$^{-1}$ region, see Figure 1c. Our calculations of methylated BP layers reproduce Raman bands below 300 cm$^{-1}$. Those intensities and positions critically depend both on the degree of functionalization and presence of counter ions, however. As an example, the calculated Raman spectrum of a methylated BP monolayer is shown in Fig 1c.

One could have expected that the electrophilic trapping with hexyliodide will lead to the formation of 2D-BP phosphonium sites incorporated in an otherwise intact hexagonal P-lattice, with a formal positive charge in the alkylated P-atoms. However, our DFT calculations points towards a P–P bond cleavage (Scheme 2). Indeed, we only obtain data for the phosphonium if a whole positive charge is included on the BP sheet. In this scenario we find a Raman fingerprint at 315 cm$^{-1}$ (SI 6), which is absent in the experimental data. To further check the suitability of the calculations, we also developed the functionalization reaction with methyliodide, showing a rather good agreement with the predictions (Figure SI 5&8). According to our calculations, we assume that the occurrence of these distinct peaks is related to P–P vibrations originating from lattice distortions in the BP sheet (see Supporting Information Videos).

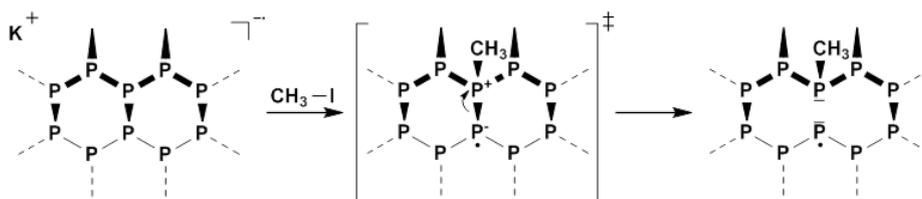

**Scheme 2.** Lattice opening of BP upon covalent modification with methyliodide: proposed reaction mechanism based on DFT calculations.

Moreover, the peaks at around 195 cm$^{-1}$ and 230 cm$^{-1}$ can be associated to turbostratic disordering of individual layers or edge phonons in the BP lattice and can even be measured in pristine BP.[31,13] We have checked the suitability of the reaction using both Na- and K-BPICs, and methyl- and hexyliodide. Mean Raman spectra (> 50 single point spectra) repeatedly exhibit the same behavior with clear bands at *ca*. 145, 217, and 308 cm$^{-1}$, thus confirming the successful functionalization reaction and the homogeneity of the samples (SI 8&11). The influence of the excitation wavelength (533 and 633 nm) shows minor changes in the relative intensity of the new modes (SI 9). The crystallinity of the methyl-functionalized specimen was also investigated via X-Ray diffraction (for experimental details see SI 12). The results can be described by a model with a few separated planes, where the symmetry along the b-axis –the crystal structure of BP is made up of puckered layers of atoms stacked along the crystallographic b-axis– has been lost, thus confirming the results stated by Raman spectroscopy.

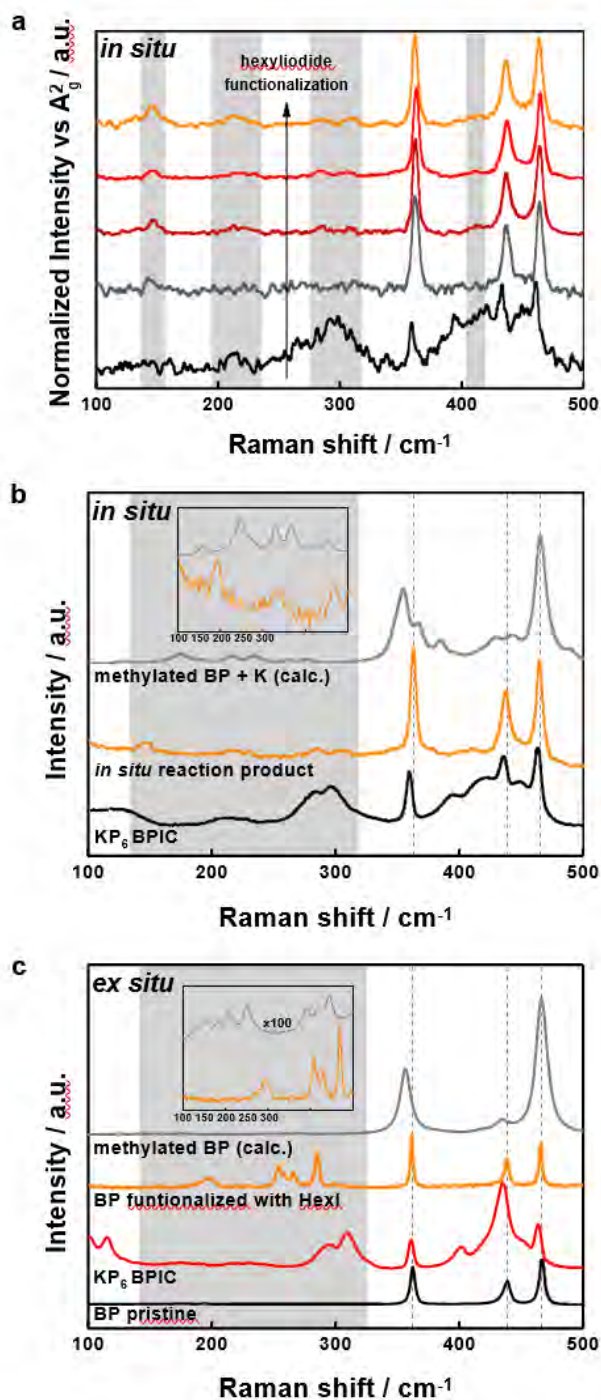

**Figure 1.** a) *In situ* Raman spectroscopy monitoring the reaction of hexyliodide with KP6 using an excitation wavelength of $\lambda_{exc}$ = 633 nm. With increasing amount of hexyliodide distinct new Raman peaks arise at 145, 210, between 260 and 285 and at 410 cm$^{-1}$. b) Calculated Raman spectrum of a methylated BP single layer saturated with potassium. The inset shows a magnification of the region below 300 cm$^{-1}$ for better comparison. The calculated spectrum has been shifted by 15 cm$^{-1}$. c) Mean Raman spectra visualizing he reaction course of the covalent functionalization of BP *ex situ*. The calculated Raman spectrum of a BP single layer with one added methyl group is also included. The inset magnifies the region below 300 cm$^{-1}$ for better comparison. The calculated spectrum has been shifted by 14 cm$^{-1}$.

In order to gain more information about the chemical composition of the covalently attached addends, TG-MS measurements were conducted. Figure 2a and SI 13 depicts the mass loss of the covalently modified BP sample upon heating up to 600 °C, which contains three interesting features. First, two mass losses between 100 °C and 300 °C which can be attributed to the detachment of the hexyl chain from the BP lattice (*m/z* = 85 and its typically related mass fragments *m/z* = 56 and *m/z* = 41). The differences in the temperatures may be related to different binding sites: basal plane atoms (zig-zag & armchair), edge near positions or rims.[32] Secondly, the sharp mass loss of the sample above 400 °C can be explained by the complete decomposition of BP to $P_4$ and $P_4$-based clusters (*m/z* = 31, 62, 93 and 124). Indeed, compared to pristine BP prepared under the same conditions but without functionalization reaction, the degradation temperature shifts from *ca*. 330 to 360 ºC, highlighting the increase of the thermal stability (SI 14). Moreover, neither hexyliodide (*m/z* = 212), $I_2$ (*m/z* = 254), nor its monoatomic equivalent (*m/z* = 127) have been detected (SI 13). This observation points towards the formation of the side product KI between the leaving group iodide and the K, which has a melting point of 723 °C and thus cannot be detected in the measurement. As a matter of fact, the residual mass loss in the functionalized sample is *ca*. 30%, which could be related to the presence of KI due to its poor solubility in THF during the work-up (*vide infra*), as well as the thermal formation of graphitic carbon from the addends (SI 15). These assumptions can be confirmed by the solubilization of KI in an aqueous solution of silver nitrate as well as the presence of benzene mass fragments, typically associated to nanocarbon degradation.[30] Finally, the abrupt mass loss at around 480 ºC may be related with trapped alkyl moieties between the layers, the formation of P-based nanoforms like nanoribbons or cluster-like species.

In order to demonstrate the covalent binding of the alkyl chains and investigate the reversibility of the functionalization reaction we performed a temperature-dependent SRS analysis. We conducted Raman mappings of the functionalized flakes deposited on Si/$SiO_2$ wafers increasing the temperature from 20 up to 220 °C (Figure 2b and SI 16). The three main modes of BP can be clearly seen as well as the characteristic peaks below 300 cm$^{-1}$ related to the covalent functionalization, which are highlighted in grey. Upon heating, the intensity of these modes starts to decrease at temperatures above 160 °C, until they nearly vanish at 200 °C in agreement with the TGA-MS analysis, indicating the de-functionalization of the BP lattice. Here, it should be mentioned that also the $A_g$, $B_{2g}$ and $A_g$ modes decrease in absolute intensity with increasing temperature, which is explained by the thermal instability of FL-BP.[14]

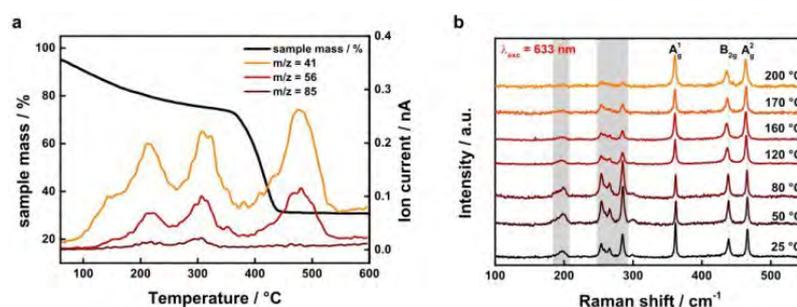

**Figure 2.** a) TG-MS displays a significant mass loss below 200 °C which can be correlated to the detachment of the covalently bound hexyl-groups before the BP lattice decomposes to P4 above 400 °C. The MS data shows characteristic mass fragments of the hexyl groups with *m/z* = 85, 56 and 41. b) T-dependent Raman spectroscopy. The disappearance of the Raman modes below 300 cm$^{-1}$ at temperatures above 170 °C can be attributed to de-functionalization of BP demonstrating the reversibility of the reaction.

Compared to other recent reports about the neutral covalent functionalization of BP using diazonium salts[20] or nucleophilic molecules[21], in which no significant change in the Raman fingerprint was observed, our results indicate a higher degree of functionalization. We also carried out the functionalization of BP with hexyliodide without alkali metal intercalation using exactly the same experimental settings. Note that for the neutral route, a previous step of liquid phase exfoliation is required prior to the functionalization. Additionally, we explored different reaction temperatures and solvents (THF and NMP), as well as inert and ambient conditions (see SI 17). Related Raman spectra show no additional vibrational modes besides the $B_{3g}$ and $B_{1g}$ modes already mentioned at 194 and 230 cm$^{-1}$. Investigations with X- Ray diffraction were performed similarly to the ones in methyl-functionalized BP. Functionalized with hexyliodide, BP exhibits a higher order and probably features a few-layer phase (as methylated BP) and a compressed phase with a three- dimensional order (see SI 12). Also, in TGA-MS analysis, no mass loss around 200 °C and no detection of characteristic mass fragments were observed (SI 18). We also conducted XPS measurements obtaining a direct proof for the formation of a P– C bond. In Figure 3 (see SI 19 for the related survey spectrum), the P 2p region of pristine BP and of the covalently modified analogue is shown. The fit comprises three spin-orbit split doublets (splitting of 0.78 eV) assigned to pristine BP P$^0$ (black), oxidized phosphorous P$^{P-O}$ (red), and carbon-bound phosphorus P$^{P-C}$ (orange). The P $2p_{3/2}$ signal of the P$^0$ species was set to 130.1 eV, which is the value determined from the measurement of the pristine BP sample, and is in agreement with a value of 130.06 eV reported for crystalline BP.[8] For the analysis of the samples, the chemical shift between pristine and oxidized phosphorous is constrained to be 4.1 eV, in accordance to the pristine sample. The analysis shows that for the hexyl- functionalized sample a carbon-bound species with a chemical shift of +2.7 eV relative to the P$^0$ signal is observed. This gives further experimental indication for the successful covalent modification of the BP lattice.

Concerning stability, we have evaluated the final functionalized flakes versus exposure to oxygen and moisture by measuring the SRM immediately after functionalization and after 15 days, showing the typical exponential decay of the Raman modes, and reflecting its degradation (SI 20). Furthermore, we have evaluated with XPS the influence of water on the covalent functionalization procedure. For this, we have submitted the sample to a final aqueous washing step during the work-up process. Interestingly, the spin-orbit split doublets assigned to the formation of P–C bonds dramatically decreases (SI 21)., while the peaks associated to potassium and iodine almost completely disappeared (data not shown), due to the high solubility of the corresponding ions in water These results point towards a water-assisted de-functionalization of the BP lattice.

Last but not least, to gain unambiguous evidence on the covalent bond formation we conducted quantitative magic angle spinning $^{31}$P solid nuclear magnetic resonance ($^{31}$P MAS NMR). Figure 4A depicts the spectra of intercalated KP$_6$, clearly showing the expected signal for phosphorene at 18.2 ppm plus a single new signal at -117 ppm, which can be assigned to axially coordinated P atoms bearing a localized negative charge and corresponds to ≈7% of the total P atoms.[33] These highfield shifted P atoms nicely agree to the proposed P atoms popping out from the 2D framework after receiving electronic density from K in the BPIC structure (Figure SI 3). Following this, the $^{31}$P MAS NMR spectrum of the methylated BP shows the complete disappearance of the negatively charged P atoms, and the appearance of a new signal at 22 ppm, a value that fits very precisely to that expected for a P–CH$_3$ and not for a P$^+$–CH$_3$ bond, and that integrates for ≈7% of the total P atoms after deconvolution (Figure 4B). Moreover, $^{13}$C MAS NMR also supports the formation of new P–CH$_3$ bonds (and not the corresponding phosphonium species).

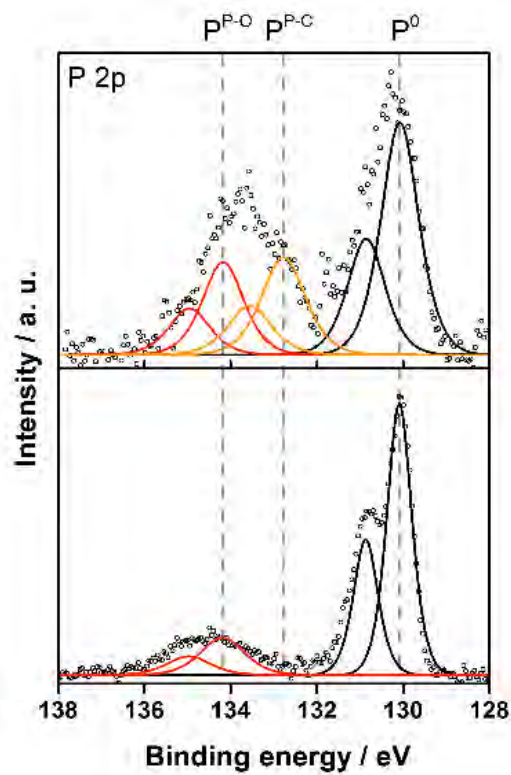

**Figure 3.** XP spectra of covalently functionalized BP with hexyliodide (top) compared to pristine BP (bottom), displaying the P 2p region, with the 2p$_{1/2}$ and 2 p$_{3/2}$ components of the doublets separated by 0.78 eV; the positions of the 2 p$_{3/2}$ levels are indicated by vertical dashed lines. The fits of the P$^0$, P$^{P-C}$ and the oxidized P$^{P-O}$ species are shown by black, orange and red lines, respectively.

In order to further confirm the functionalization, the spectrum was acquired in $^1$H–$^{31}$P cross-polarization mode, such that only the P atoms having H atoms at 1 or 2 atom bond distance will be detected (Figure 4B). The corresponding spectra shows the persistence of the signal at 22 ppm and the complete disappearance of the original phosphorene signal at 18.2 ppm, which strongly supports the formation of a covalent P–CH$_3$ bond. Together, these NMR results indicate that the KP$_6$BPIC possesses a ≈7% of localized and negatively charged P atoms that react quantitatively in a substitution reaction with methyliodide to give new covalent P–CH$_3$ bonds and KI, in stark contrast to the neutral route (SI 22). This covalent functionalization presents better values than that of related alkyl- functionalized graphene and, in terms of solubility, a similar behaviour in o-DCB, indicative of improved processability (SI 23).[30]

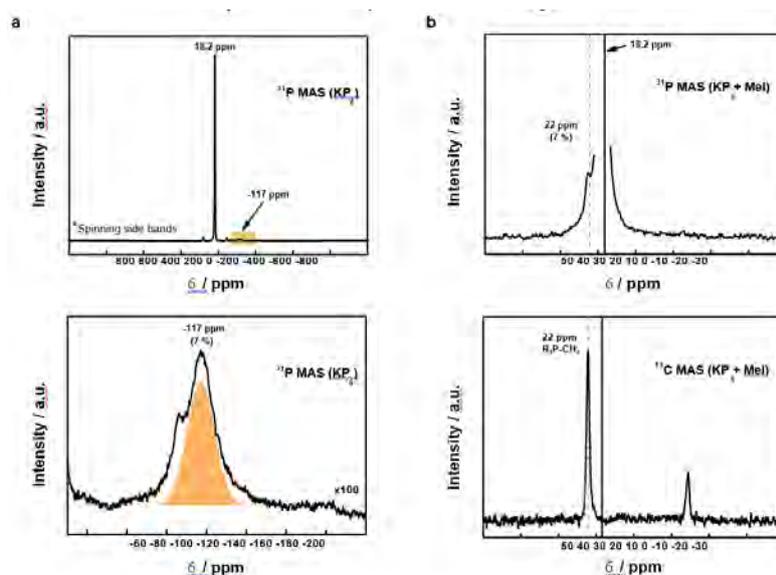

**Figure 4.** a) $^{31}$P MAS NMR spectra of intercalated BP (KP$_6$) featuring the signal for single black phosphorus at 18.2 ppm as well as a signal at -117 ppm, which can be assigned to axially coordinated P atoms bearing a localized negative charge. b) (Top) $^{31}$P MAS NMR spectrum of BP functionalized with methyl moieties showing the appearance of a new signal at 22 ppm confirming the presence of P-CH$_3$ species. (Bottom) Accordingly, the $^{13}$C MAS NMR spectrum acquired in $^1$H-$^{31}$P cross-polarization mode shows the disappearance of the original BP signal at 18.2 ppm and the persistence of the signal at 22 ppm, strongly supporting the formation of a covalent P–CH$_3$ bond.

In conclusion, the reductive covalent functionalization of BP with alkyl halides using intercalation compounds result in a remarkably high degree of functionalization and involves a P–P bond breakage (Scheme 2). These results open new pathways for the fine-tuning of the BP chemical and physical properties, as well as the development of unprecedented hybrid materials.

**Acknowledgements**

A.H. and G.A. acknowledge the European Research Council (ERC Advanced Grant 742145 B-PhosphoChem to A.H., and ERC Starting Grant 2D- PnictoChem 804110 to G.A.) for support. The research leading to these results was partially funded by the European Union Seventh Framework Programme under grant agreement No. 604391 Graphene Flagship. G.A. has received financial support through the Postdoctoral Junior Leader Fellowship Programme from "la Caixa" Banking Foundation (LCF/BQ/PI18/11630018).

G.A. thanks support by the Deutsche Forschungsgemeinschaft (DFG; FLAG- ERA AB694/2-1), the Generalitat Valenciana (SEJI/2018/034 grant) and the FAU (Emerging Talents Initiative grant #WS16-17_Nat_04). A.H. and A.G. thank the SFB 953 "Synthetic Carbon Allotropes" funded by the DFG for support and the Cluster of Excellence „Engineering of Advanced Materials".

A.M. thanks Alexander von Humboldt (AvH) Foundation for a postdoctoral fellowship. This work was supported by the MINECO (Spain) through the Excellence Unit María de Maeztu (MDM-2015-0538) and the Project CTQ2017–86735–P.

This work was supported by the Deutsche Forschungsgemeinschaft (DFG) through the SFB 953, the Alexander von Humboldt (AvH) Foundation (Tao Wei) and Marie Sklodowska-Curie IF European Actions 747734 *Hy-solFullGraph* to M.E. Pérez-Ojeda.

# Supporting Information

## Lattice Opening Upon Bulk Reductive Covalent Functionalization of Black Phosphorus

Stefan Wild, Michael Fickert, Aleksandra Mitrovic, Vicent Lloret, Christian Neiss, José Alejandro Vidal-Moya, Miguel Ángel Rivero-Crespo, Antonio Leyva-Pérez, Katharina Werbach, Herwig Peterlik, Mathias Grabau, Haiko Wittkämper, Christian Papp, Hans-Peter Steinrück, Thomas Pichler, Andreas Görling, Frank Hauke, Gonzalo Abellán* and Andreas Hirsch*

[a]     S. Wild, M. Fickert, Dr. A. Mitrovic, V. Lloret, Dr. F. Hauke, Dr. G. Abellán, Prof. A. Hirsch
Chair of Organic Chemistry II and Joint Institute of Advanced Materials and Processes (ZMP); Friedrich-Alexander-Universität Erlangen-Nürnberg (FAU); Nikolaus-Fiebiger Straße 10, 91058 Erlangen and Dr.-Mack Straße 81, 90762 Fürth (Germany)

E-mail: andreas.hirsch@fau.de and gonzalo.abellan@fau.de

[b]     Dr. C. Neiss, Prof. A. Görling
Lehrstuhl für Theoretische Chemie and Interdisciplinary Center of Molecular Materials (ICMM); Friedrich-Alexander-Universität Erlangen-Nürnberg (FAU); Egerlandstraße 3, 91058 Erlangen (Germany)

[c]     Dr. J. A. Vidal-Moya, M. A. Rivero-Crespo, Dr. A. Leyva-Pérez
Instituto de Tecnología Química. Universidad Politècnica de València–Consejo Superior de Investigaciones Científicas. Avda. de los Naranjos s/n, 46022, Valencia (Spain)

[d]     K. Werbach, Prof. H. Peterlik, Prof. T. Pichler
Faculty of Physics, University of Vienna; Strudlhofgasse 4, 1090 Vienna (Austria)

[e]     M. Grabau, H. Wittkämper, Dr. C. Papp, Prof. Dr. H.-P. Steinrück
Lehrstuhl für Physikalische Chemie II, FAU, Egerlandstraße 3, 91058 Erlangen (Germany)

[f]     Dr. G. Abellán,
Instituto de Ciencia Molecular (ICMol), Universidad de Valencia,
Catedrático José Beltrán 2, 46980, Paterna, Valencia (Spain)

## Experimental Section:

### Materials and functionalization process:

For all experiments BP purchased from Smart Elements with purity higher than 99.999% was used. First, BP crystals were mortared before they were intercalated with potassium/ sodium to create $KP_6$ or $NaP_6$ following the procedure, which was previously described by our own working group.[1] The obtained Black Phosphorus Intercalation Compounds (BPICs) were dispersed in purified THF (1 mg/ml). In order to separate the activated BP the dispersion was treated by ultra-sonication using a Bandelin SONOPULS HD4100 sonotrode for 5 minutes (power amplitude: 25%; time interval: 2 s) yielding a more stable dispersion of negatively charged BP layers. In the next step, hexyl iodide or methyl iodide (Sigma Aldrich) was added as electrophilic functionalization reagent (1 eq. per phosphorous atom) and the dispersion was stirred for 1 hour, before the functionalized BP powder was obtained by filtration. The whole functionalization process was performed in an argon filled LABmaster[pro] sp glove box (MBraun) equipped with a gas purifier and solvent vapor removal unit (oxygen and water content lower than 0.1 ppm).

### Solvent purification:

THF was first dried over Molecular Sieves with a size of 3 Å for at least 3 days to remove dissolved water. With this step, the water content could be determined by means of the "Karl Fischer" method to be lower than 5 ppm. Afterwards, the THF was degassed using the "Freeze-Pump-Thaw" method to remove oxygen, before it was introduced into the glove box. Additionally, the THF was distilled in the glove box over a Na/K alloy to further reduce the water content.

### Characterization:

*In situ Raman Spectroscopy*: *In situ* Raman spectroscopic detection was carried out inside a quartz tube through a flat (0.7 mm thick) optical window of borosilicate glass (PGO GmbH) in an ultra-high vacuum chamber at ca. $3 \times 10^{-7}$ mbar where the BPIC ($KP_6$ or $NaP_6$) was placed in a sample boat and the hexyl iodide as trapping reagent in a vial connected to separate control unit for the pressure. The Raman measurements were performed at room temperature using a Horiba LabRam spectrometer with a 633 nm excitation wavelength at 0.5 mW between 300 and 3000 cm$^{-1}$.

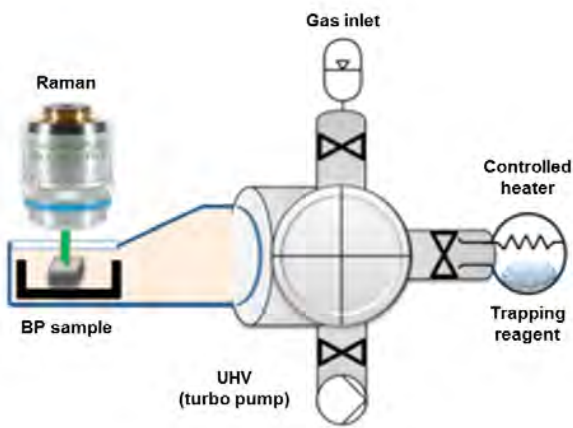

**Figure SI 1:** Schematic illustration of the setup for the controlled reaction of a BPIC with hexyl iodide as functionalization reagent under ultra-high vacuum conditions. The reaction progress is monitored by *in situ* Raman spectroscopy.

*Raman Spectroscopy*: Raman spectra were acquired using a LabRam HR Evolution confocal Raman microscope (Horiba) equipped with an automated XYZ table using 0.80 NA objectives. An excitation wavelength of 532 nm or 633 nm was used preferentially to characterize the covalently functionalized BP. For better resolution of the spectra, a grating of 1800 grooves/mm was selected, whereat an acquisition time of 2 s was used. Additionally, the laser intensity was kept below 5% (0.88 mW) to avoid photo-induced laser oxidation of the samples. Temperature-dependent Raman studies were conducted in a Linkam stage THMS 600 equipped with a liquid nitrogen pump TMS94 for temperature stabilization under a constant flow of nitrogen The heating rate was set at 10 K·min$^{-1}$. The measurements were carried out using an excitation wavelength of 633 nm for better comparison to the results, which have been obtained by the *in situ* Raman experiments. For all conventional Raman studies the samples were drop casted on a Si/SiO$_2$ (300 nm oxide layer) substrate.

*Thermogravimetric analysis coupled to Mass Spectrometry*: Thermogravimetric analysis was performed on a Netzsch STA 409CD Skimmer equipped with an EI ion source and a quadrupole mass spectrometer. Time-dependent temperature profile: 25–600 °C (10 K min$^{-1}$ temperature ramp) and cooling to 30 ºC. The initial sample weights were adjusted at 5.0 (±0.1) mg and the whole experiment was executed under inert gas atmosphere with a He gas flow of 80 mL min$^{-1}$.

*X-Ray Photoelectron Spectroscopy*: X-ray photoemission (XP) spectra of covalently functionalized and pristine BP were collected at room temperature. Synthesis, preparation and transfer into the XPS chamber took place in Argon, i.e. in oxygen-free atmosphere. Drop-cast preparation onto unpolished gold foil, which was sonicated in isopropanol for at least 2 min and dried in air before transfer to Argon atmosphere. The shown data was collected using monochromatized Al K$_\alpha$ radiation. The base pressure of the UHV system was within the low 10$^{-10}$ mbar regime. The pass energy was 10 eV for the acquisition of narrow, high-resolution scans of certain regions of the spectrum, and 50 eV for the acquisition of survey spectra. Unless stated differently, the given binding energies are referenced to the Fermi energy of the Gold support material, and calibration was done using the binding energy of Au 4f$_{7/2}$ signals, which was set to be 84.0 eV in accordance to literature. Linear baselines were adjusted to narrow P$_{2p}$ region scans. For the analysis pseudo-Voigt profiles, that is a product of weighted Lorentzian and Gaussian with 40 % Gaussian contribution, were fitted to the data.

*X-Ray Diffraction*: XRD-experiments were performed with Cu-K$_\alpha$ radiation (wavelength 0.1542 nm) from a microfocus source (Incoatec High Brilliance), equipped with a pinhole camera system (Nanostar, Bruker AXS), a 2D position sensitive detector (Vantec 2000) and an image plate system (Fuji). The samples were transported in sealed capillaries from the glove box to the measurement chamber of the X-ray equipment and measured in vacuum. The intensity data were radially integrated and background corrected to result in intensities in dependence on the scattering angle 2θ in a range from about 10 to 70 degrees.

*Density Functional Theory Calculations*: DFT calculations were carried out using the Vienna ab-inito Simulation Package (VASP 5.4) employing a plane wave basis set within the PAW approach[2-5] and the exchange-correlation functional according to Perdew, Burke, Ernzerhof (PBE).[6] Van-der-Waals bonding effects were captured by the dispersion correction "D3" by Grimme (without Becke-Johnson damping).[7-8] Atomic cores were modelled by the PAW potentials of VASP 5.4, leaving five / four / one / nine / seven electrons in the valence region of P / C / H / K / I atoms, respectively. The plane wave cut-off energy was set to 450 eV throughout (except for Raman intensities, see below).

Geometry optimizations were done with the help of an external driver "Gadget" by Bučko et al.,[9] and were considered converged if all cartesian force components are below 0.005 eV/Å. All calculated systems were checked for spin polarization and re-optimized if necessary. Methfessel-Paxton smearing of first order with $\sigma=0.1$ eV was used.[10] To sample the first Brillouin zone Monkhorst-Pack meshes (always shifted to include the $\Gamma$-point) were employed.[11] The number of k-points was chosen such that a k-point distance of 0.02 – 0.01 $2\pi$/Å per reciprocal direction was obtained, which leads, for example, in case of the 4x3 super cell of a BP single layer to a 4x4x1 k-point mesh.

Vibrational frequencies and phonon modes were calculated at the $\Gamma$-point within the harmonic approximation. Raman intensities of the modes were calculated via density-functional perturbation theory (DFPT) employing a larger cut-off energy of 500 eV, Gaussian smearing with $\sigma=0.001$ eV, and a very tight SCF convergence criterion of $10^{-8}$ eV. PAW projections were done in reciprocal space, which is more accurate.

In the course of the study several theoretical models were considered to elucidate the effects of charging the BP layers, the presence of counter ions, and the effects of degree of functionalization. In more detail, the following systems were treated theoretically (in parenthesis we state whether the final system is spin polarized or not):

- Pristine single-layer BP, primitive unit cell containing four P atoms (non-spin polarized)
- Methylated 4x3 single-layer BP (spin polarized)
- Methylated 4x3 single-layer BP, positively charged with a compensating background charge (non-spin polarized)
- Methylated 4x3 single-layer BP with a saturating K atom (non-spin polarized)
- Methylated 4x3 single-layer BP with a saturating K atom, positively charged with a compensating background charge (spin polarized)
- Methylated 4x3 single-layer BP with a saturating I atom (non-spin polarized)
- Methylated 4x3 single-layer BP with a saturating I atom, negatively charged with a compensating background charge (non-spin polarized)
- Methylated 4x3 single-layer BP with saturating K and I atoms (spin-polarized)
- Methylated 2x2 single-layer BP (spin polarized)
- Methylated 1x1 single-layer BP (non-spin polarized)

Each of the models were subject to full structural relaxation and subsequent vibrational analysis and calculation of Raman intensities.

We found that both charging the methylated 4x3 single-layer BP with a saturating K atom, and saturating the methylated 4x3 single-layer BP with K-I resembles the structure and properties of the methylated 4x3 single-layer BP without additional atoms (i.e., the P-P bond next to the functionalization site is significantly elongated – as described in the main article – which is accompanied by an unpaired electron). This resemblance is also true for the Raman spectra in these cases which are very similar in these cases if one considers the positions of the Raman peaks, see Fig. SI 22.

In all cases, except the positively charged methylated 4x3 single-layer BP, one P-P bond is clearly elongated, see Figs. SI2, SI3. Potassium or iodine quench the unpaired spin by saturating the dangling bond.

In case of the positively charged methylated 4x3 single-layer BP we find in contrast a phosphonium-like structure of the functionalized P-atom, see Fig. SI 23, which is also responsible for a clear peak in the Raman spectrum at 314 cm$^{-1}$ (calculated value), see Fig. SI 23. This peak might be viewed as a fingerprint of the phosphonium-like motif.

## SI 2: Structure and spin density of a methylated BP single layer

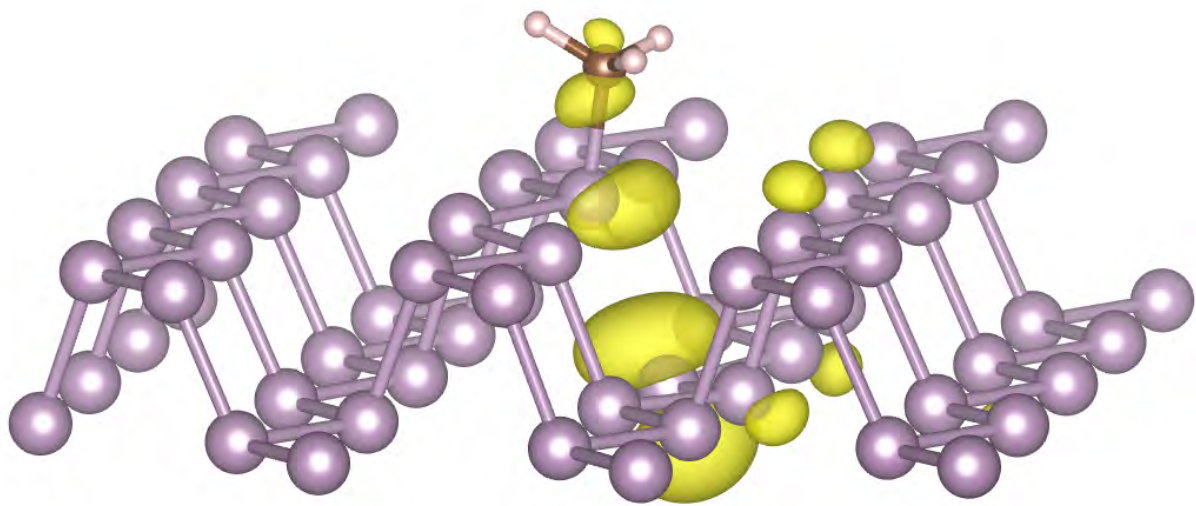

**Figure SI 2:**
Structure and spin density of a methylated BP single layer displaying a 4 x 3 super cell of BP and one added methyl group. Full geometry relaxation (incl. cell parameters) was allowed. The cell parameters were found to change only very little.

## SI 3: Saturation of a methylated BP single layer with iodine and potassium

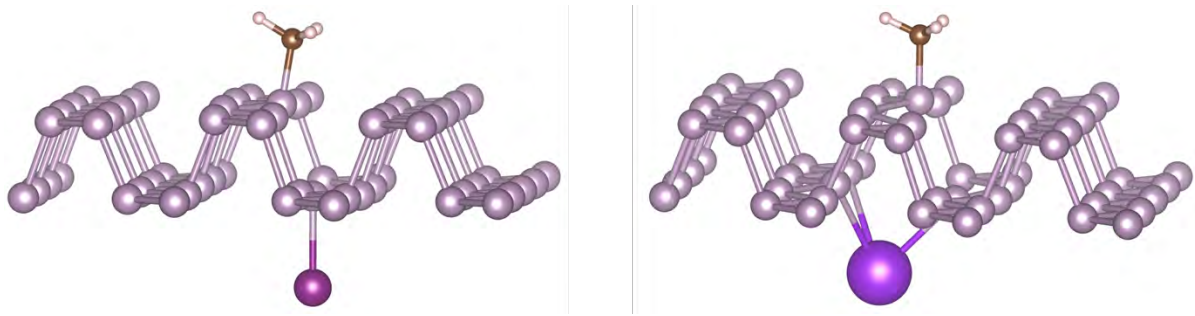

**Figure SI 3:**
Structure of a methylated BP single layer saturated with iodine (left) and potassium (right): Saturation removes the radical character of the functionalized BP layer and restores a band gap of 0.95 eV (iodine) or 0.50 eV (potassium), respectively. At this point, it is worth to mention that saturation with a second alkyl group would also be possible, but then reduction of iodine would be expected, which was not observed experimentally.

**SI 4: Calculated Raman spectra of a methylated BP single layer: effect of counter atoms**

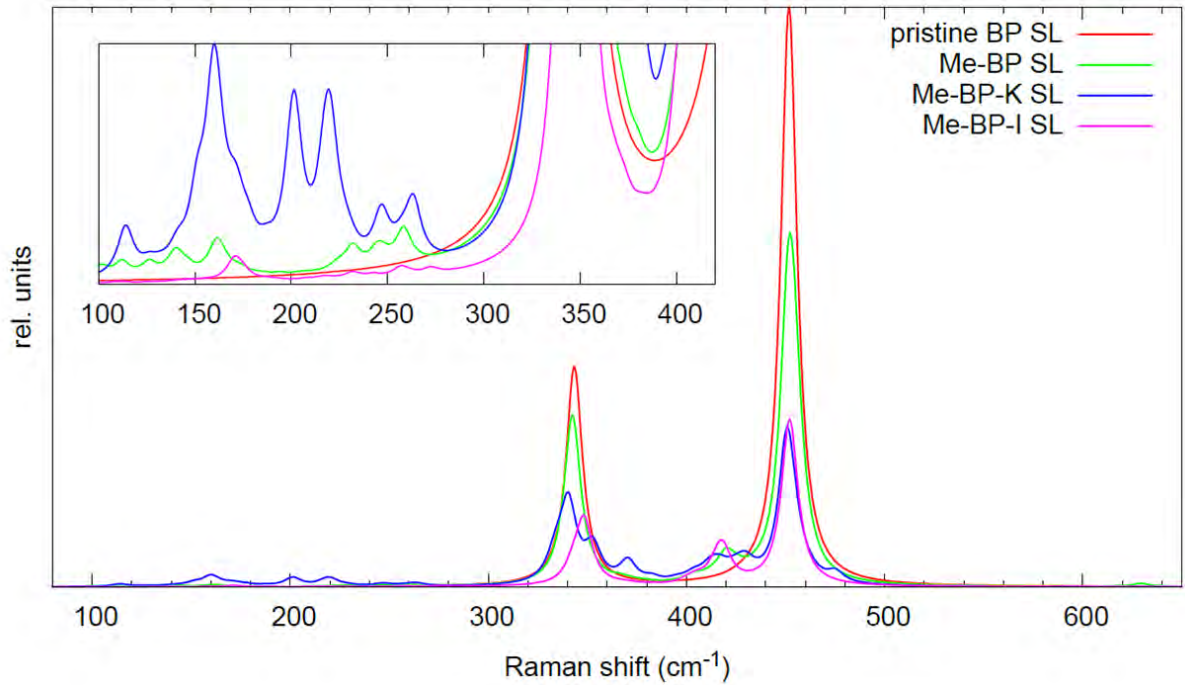

**Figure SI 4:**
Calculated Raman spectra below 650 cm$^{-1}$ of a (methylated) BP single layer (4 x 3 super cell): The inset magnifies the range between 100 cm$^{-1}$ and 420 cm$^{-1}$. A Lorentzian broadening was applied (HWHM = 5.0 cm$^{-1}$).

**SI 5: Calculated Raman spectra of a methylated BP single layer (bare, with K, with K$^+$, and KI)**

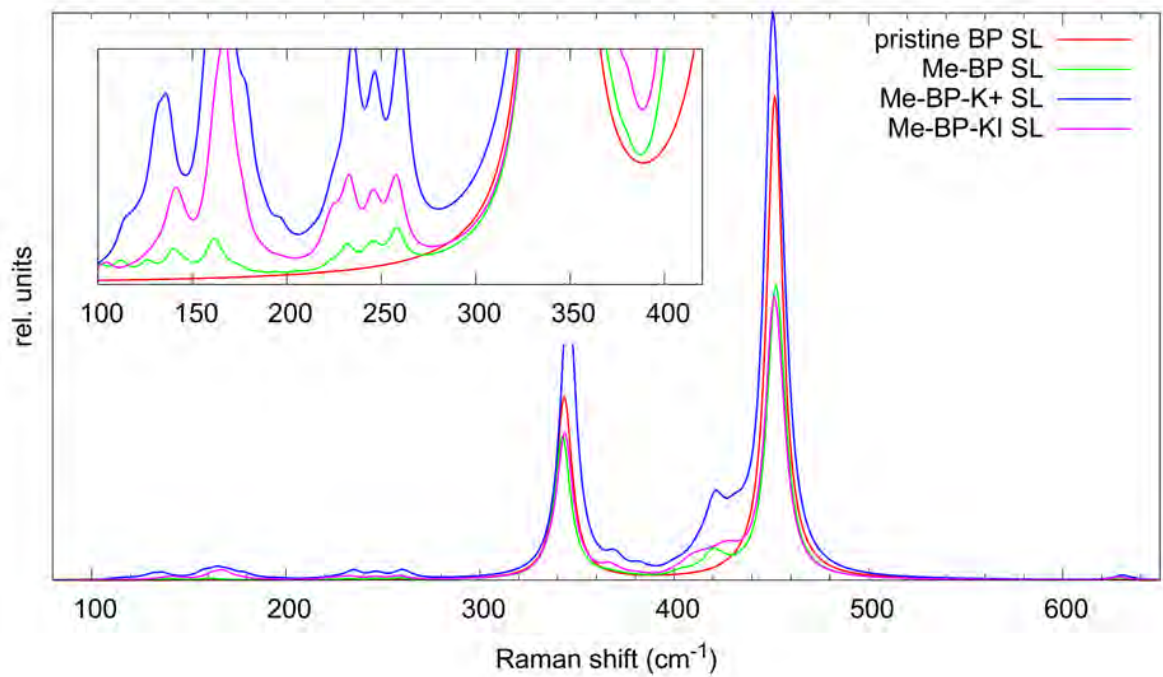

**Figure SI 5:**
Calculated Raman spectra below 650 cm$^{-1}$ of a (methylated) BP single layer (4 x 3 super cell): The inset magnifies the range between 100 cm$^{-1}$ and 420 cm$^{-1}$. A Lorentzian broadening was applied (HWHM = 5.0 cm$^{-1}$).

**SI 6: Calculated Raman spectra of a methylated BP single layer: influence of charging**

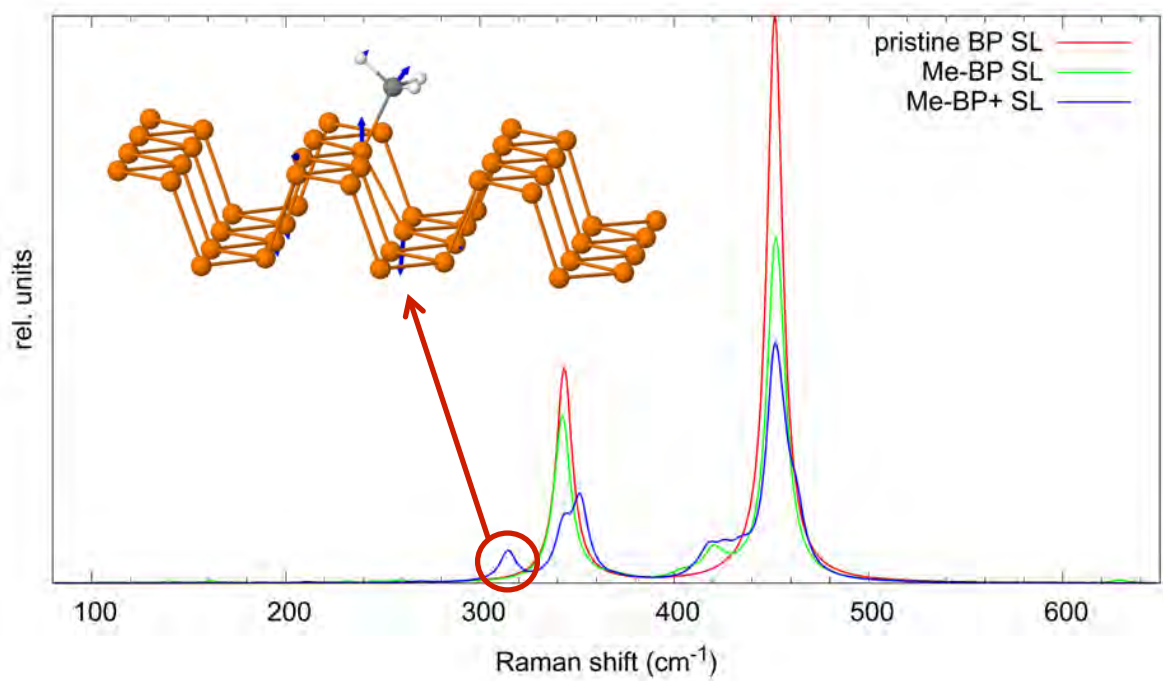

**Figure SI 6:**
Calculated Raman spectra below 650 cm$^{-1}$ of a (methylated) BP single layer (4 x 3 super cell). A Lorentzian broadening was applied (HWHM = 5.0 cm$^{-1}$).

**SI 7:** *In situ* Raman Spectroscopy of NaP$_6$ functionalized with hexyliodide

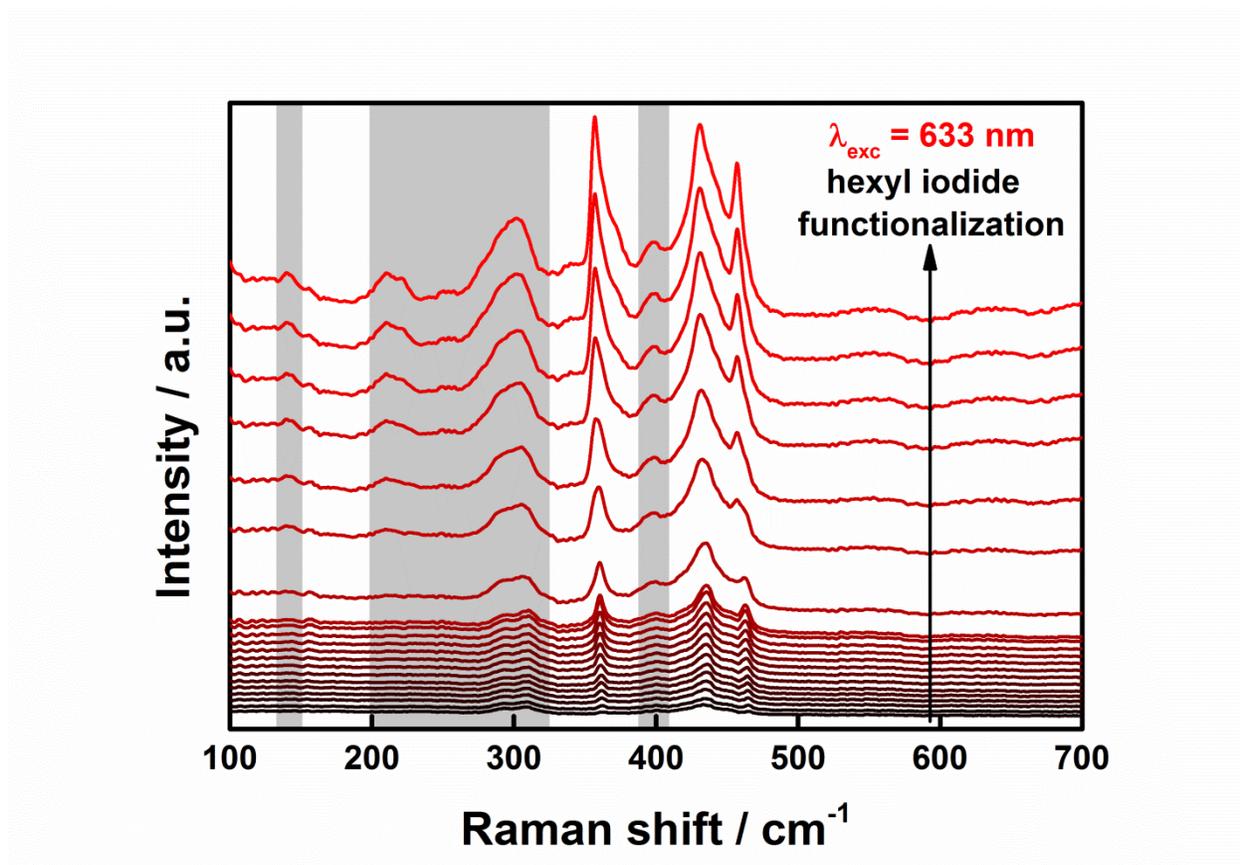

**Figure SI 7:**

*In situ* Raman Spectroscopy monitoring the reaction of hexyliodide with the BPIC NaP$_6$ using an excitation wavelength of $\lambda_{exc}$ = 633 nm. With increasing amount of the electrophile hexyliodide distinct new Raman peaks arise at 145 cm$^{-1}$, 210 cm$^{-1}$, between 260 cm$^{-1}$ and 285 cm$^{-1}$ and at around 405 cm$^{-1}$ visualizing a change in the BP lattice indicative of a covalent attachment of the trapping reagent.

**SI 8: Reductive Covalent Functionalization of $KP_6$ and $NaP_6$ with methyl iodide as trapping reagent**

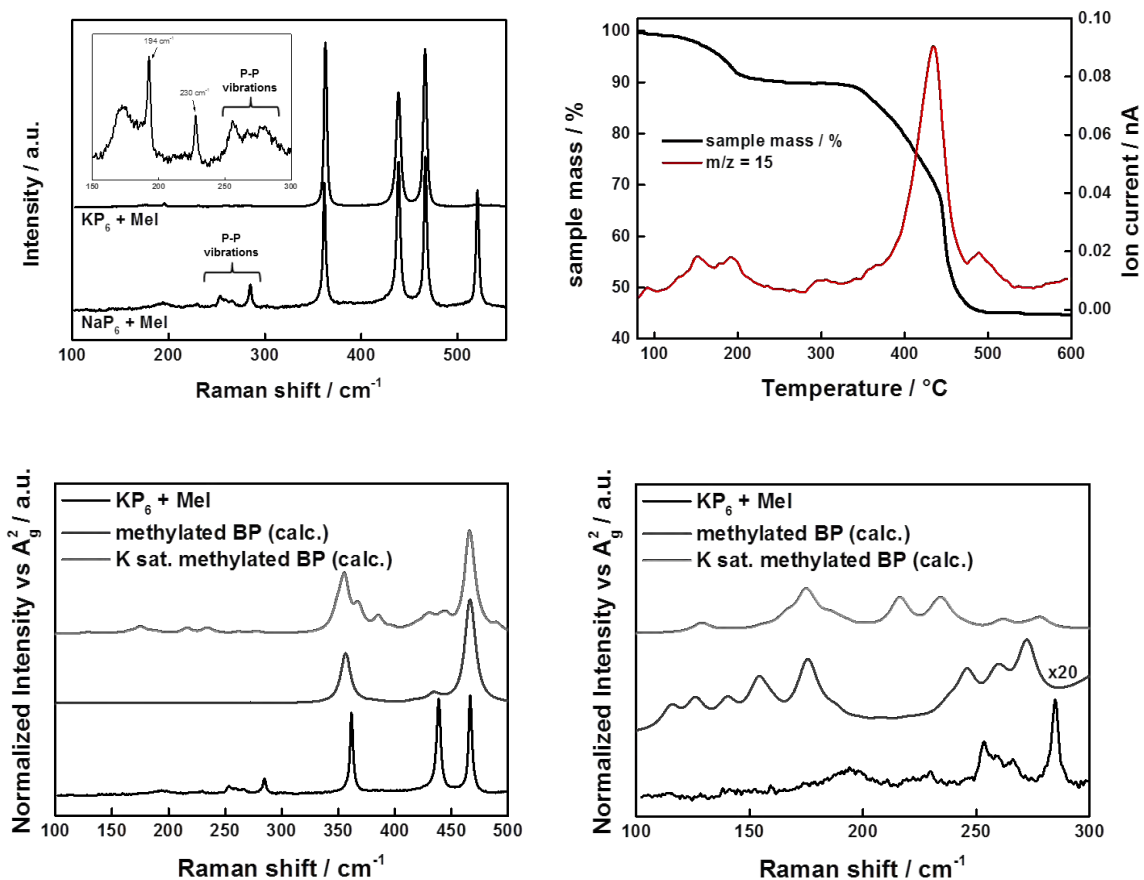

**Figure SI 8:**
(Top-Left) Mean Raman spectra of covalently functionalized $NaP_6$ (bottom) and $KP_6$ (top) with methyl iodide recorded at an excitation wavelength of 532 nm: The inset highlights the region below 300 $nm^{-1}$ of the upper spectrum for better comparison of both spectra. (Top-Right) TGA-MS analysis of $KP_6$ functionalized with methyl iodide: the detection of the methyl fragment can be attributed to the mass loss between 100 °C and 250 °C (detachment of covalently bound methyl addends) as well as to the decomposition of the BP to $P_4$ above 400 °C (release of formed phosphonium species / trapped methyl iodide between the layers). (Bottom) Comparison of experimentally obtained Mean Raman spectra of methyl functionalized $KP_6$ with DFT calculations of methylated BP and potassium saturated methylated BP. For clarity the bottom right graph shows a zoom of the low wavenumber region between 100 and 300 $cm^{-1}$.

**SI 9**: SRM of covalently functionalized BP *via* reductive route on Si/SiO$_2$

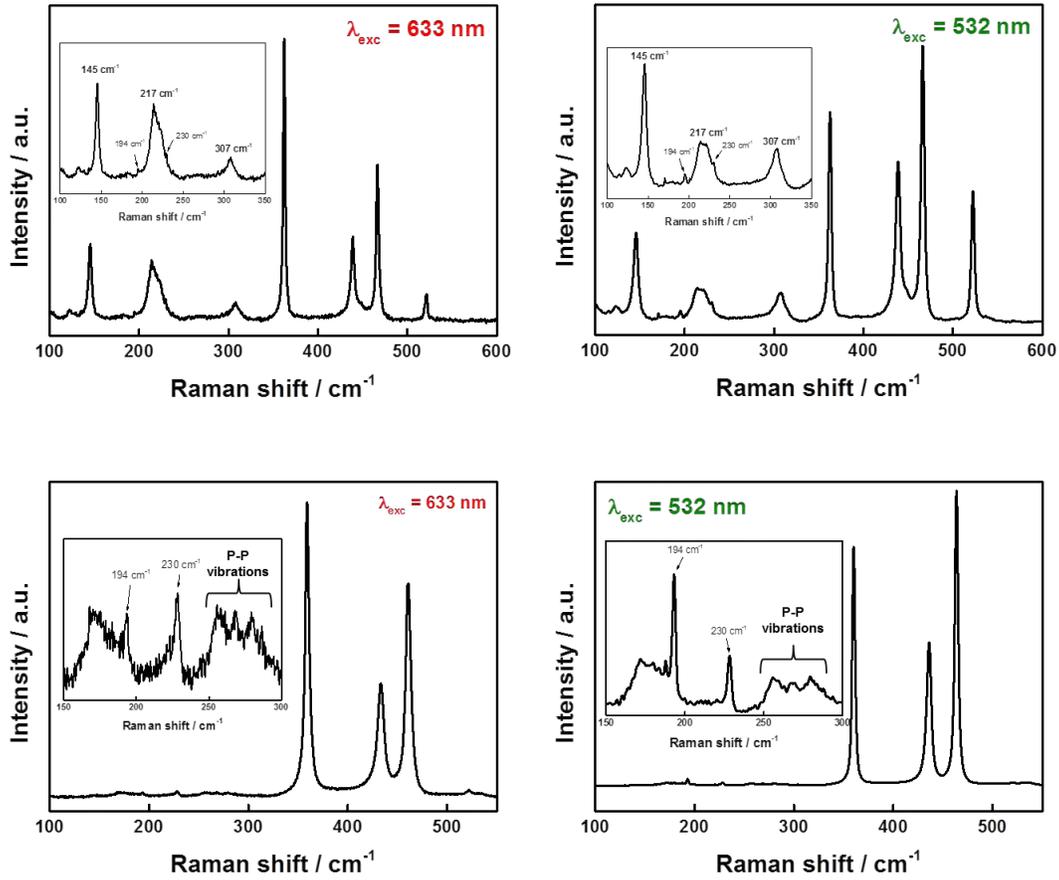

**Figure SI 9:**

Mean Raman spectra of covalently functionalized KP$_6$ with hexyl iodide recorded at different excitation wavelengths: The insets highlight region below 300 cm$^{-1}$. (Top) Mean Raman spectra showing modes similar to the ones obtained in the *in situ* experiment. This indicates that potassium is coordinated to the analyzed area. (Bottom) Mean Raman spectra of another area on the same sample displaying vibrations, which can be interpreted as P-P lattice distortions without any remaining metal (compare SI 2). The Raman modes at 194 cm$^{-1}$ and 230 cm$^{-1}$ are attributed to turbostratic disordering or edge phonons. Indeed, they seem to be more intense and can be found more frequently when the material has been functionalized than in pristine phosphorous.

**SI 10: Dependence on coverage**

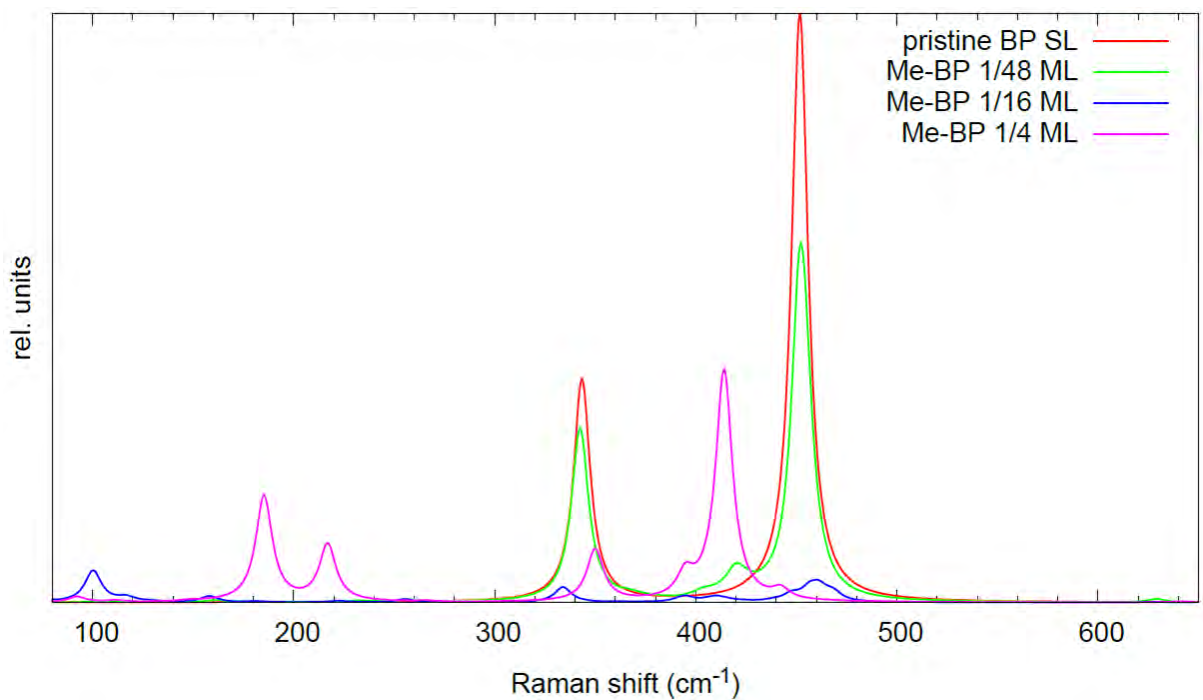

**Figure SI 10:**
Calculated Raman spectra below 650 cm$^{-1}$ of a methylated BP single layer displaying the dependence on the coverage of covalently bound addend onto the BP lattice: 1/*x* ML means one methyl group per x phosphorous atoms (single layer BP). A Lorentzian broadening was applied (HWHM = 5.0 cm$^{-1}$).

**SI 11: Reductive Covalent Functionalization of NaP$_6$ with hexyl iodide – Raman and TGA analysis**

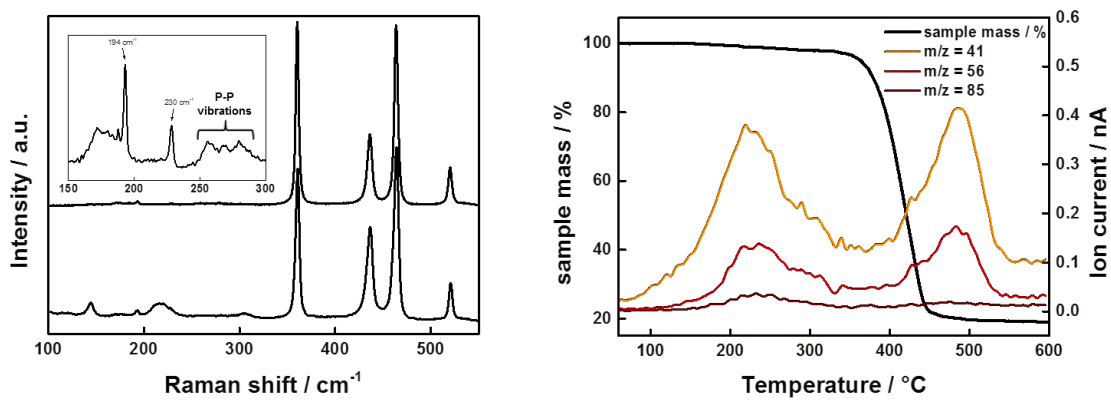

**Figure SI 11:**

(Left) Mean Raman spectra of functionalized NaP$_6$ with hexyl iodide recorded at an excitation wavelength of 532 nm: The difference in both spectra again can be explained by the coordination of remaining sodium to the BP lattice (see SI 7). Both spectra are similar to the spectra obtained when using KP$_6$ indicating that the functionalization is independent on the intercalated alkali metal. (Right) Corresponding TGA-MS analysis of NaP$_6$ functionalized with hexyl iodide: Again the detection of characteristic hexyl fragments can be attributed to the mass loss between 100 °C and 300 °C (5%; detachment of covalently bound hexyl addends) as well as to the decomposition of the BP to P$_4$ above 400 °C (release of formed P-containing species / trapped hexyl iodide between the layers).

## SI 12: X-Ray Diffraction

X-ray diffraction (XRD) experiments were performed with Cu-K$_\alpha$ radiation (wavelength 0.1542 nm) from a microfocus source (Incoatec High Brilliance), equipped with a pinhole camera system (Nanostar, Bruker AXS), a 2D position sensitive detector (Vantec 2000) and an image plate system (Fuji). The samples were transported in sealed capillaries from the glove box to the measurement chamber of the X-ray equipment and measured in vacuum. The intensity data were radially integrated and background corrected to result in intensities in dependence on the scattering angle 2θ in a range from about 10 to 70 degrees.

The samples preserved the in-plane crystallinity with a substantial decrease in the (020) and (040) layer peaks, indicative of turbostratic disorder, strongly pronounced for methylated BP and only slightly visible for hexylated BP. In comparison to pristine BP, new peaks arise in hexylated BP (SI 19). The dashed gray lines are peak positions calculated for an orthorhombic system, which is slightly compressed (lattice parameters are a=0.3308 nm, b=1.0481 nm, and c=0.4236 nm in comparison to values of pristine BP, a=0.33133 nm, b=1.0473 nm, and c=0.4374 nm).[12] The peaks for methylated BP are rather broad, which indicates a lower order in comparison to the hexyl sample, and are described by a model with separated planes (two or few layer model), where the Cmca symmetry of BP has been lost. It is proposed that the lack in symmetry in the b-axis leads to an additional reflecting plane in b-axis and the structure is described with lattice parameters a=0.3518 nm, b=0.2478 nm, and c=0.4127 nm (dashed red lines). The indices of the identified reflections are also found in the Tables below. Differently, the hexyl functionalized sample cannot be described successfully with one single lattice structure, thus two phases are used, a distorted (compressed) phase and a separated phase (two or few layer model). The two or few layer phase consists of a small number of randomly restacked nanosheets, and therefore to the presence of turbostratic disorder, as previously stated by Raman spectroscopy.

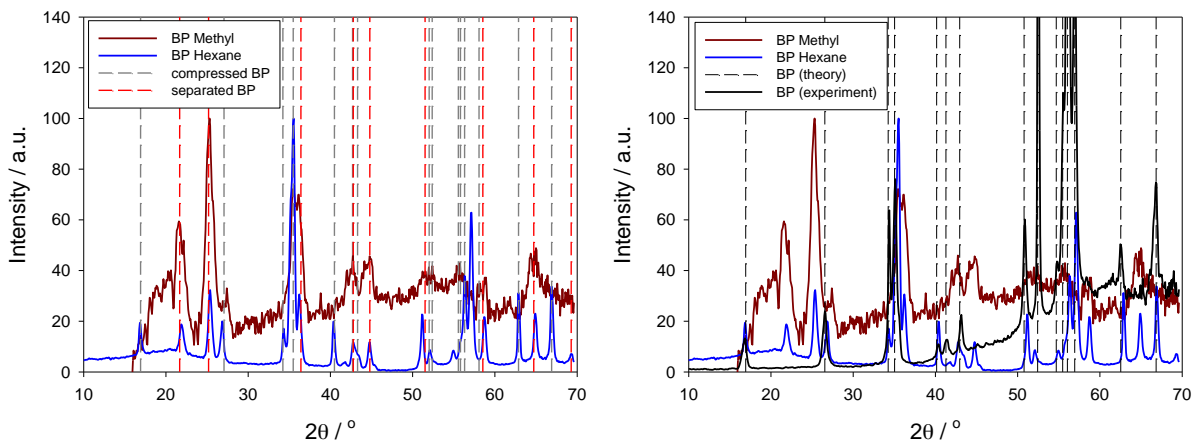

**Figure SI 12:**

(Left): X-ray intensities for BPHexyl (BP functionalized with hexyliodide, blue line) and BPMethyl (methylated BP, dark red line) in dependence on the scattering angle 2θ. The higher order for BPHexyl is clearly visible by the much less noise in the data. The dashed lines show the fit values for a slightly distorted BP phase (dashed gray line) and for a two or few layer material (dashed red line). Peaks in BPMethyl are found only for the two layer material, whereas BPHexyl exhibits both types of phases. (Right): X-ray intensities for pristine BP (black line), BPHexyl (blue line) and BPMethyl (dark red line) in dependence on the scattering angle 2θ. In comparison to pristine BP, additional peaks and a slight deviation of peak maxima due to a minimal distortion are visible for BPHexyl (blue line), whereas BPMethyl has a much higher disorder (more noisy data) and a complete absence of peaks from a layered structure (such as the (020)-reflection at a scattering angle of 16.9 degree).

**Table S1:** Experimental and calculated peaks for distorted compressed BPHex phase, orthorhombic structure with a=0.3308 nm, b=1.0481 nm, c=0.4236 nm. Values of pristine BP are a=0.33133 nm, b=1.0473 nm, c=0.4374 nm.

| Reflection | Exp BPHex 2θ / ° | Calc 2θ / ° |
|---|---|---|
| (020) | 16.85 | 16.92 |
| (021) | 26.87 | 27.07 |
| (040) | 34.32 | 34.22 |
| (111) | 35.51 | 35.48 |
| (041) | 40.39 | 40.48 |
| (002) | 42.75 | 42.69 |
| (131) | 43.5 | 43.30 |
| (112) | 52.05 | 52.01 |
| (060) |  | 52.38 |
| (200) | 54.8 | 55.56 |
| (042) | 56.3 | 55.82 |
| (132) | 57.1 | 58.06 |
| (221) | 62.8 | 62.87 |
| (240) | 66.9 | 66.90 |

**Table S2:** Experimental and calculated peaks for the separated phase (two or few-layer model), orthorhombic structure with a=0.3518 nm, b=0.2478 nm, c=0.4127 nm.

| Reflection | Exp BPHex 2θ / ° | Exp BPMeth 2θ / ° | Calc 2θ / ° |
|---|---|---|---|
| (001) | 21.88 | 21.5 | 21.67 |
| (100) | 25.35 | 25.3 | 25.18 |
| (010) | 36.23 | 36.1 | 36.42 |
| (011) | 42.75 | 42.7 | 42.78 |
| (110) | 44.75 | 44.8 | 44.79 |
| (102) | 51.1 |  | 51.52 |
| (012) | 58.7. | 59 | 58.53 |
| (112) | 64.9 | 64.4 | 64.72 |
| (221) | 69.3 |  | 69.27 |

## SI 13: TGA-MS analysis of KP$_6$ functionalized with hexyl iodide

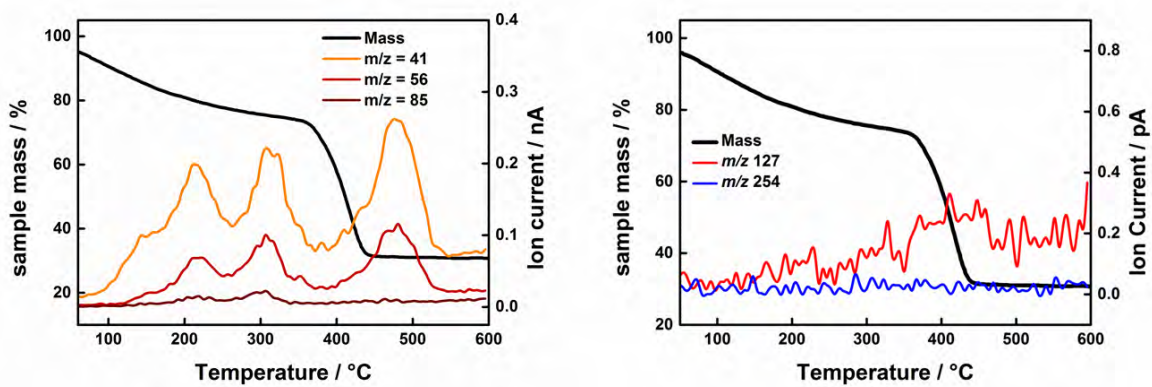

**Figure SI 13:**

(Left) TG analysis displays a significant mass loss below 200 °C which can be correlated to the detachment of the covalently bound hexyl-groups before the BP lattice decomposes to P$_4$ above 400 °C. The MS data shows mass fragments characteristic of the hexyl groups with $m/z$ = 85, $m/z$ = 56 and $m/z$ = 41. The huge detection of these mass fragments above 400 °C might originate from hexyl chains trapped between BP layers. (Right) Additionally, the signal of the ion current for the detection of iodine ($m/z$ = 254) and its monoatomic equivalent ($m/z$ = 127) are not detectable or within the noise level of the measurement, indicating that the salt KI (m.p. = 723 °C) was formed during the functionalization reaction.

## SI 14: Decomposition of BP to P$_4$: TGA-MS analysis of KP$_6$ functionalized with hexyl iodide compared to pristine BP

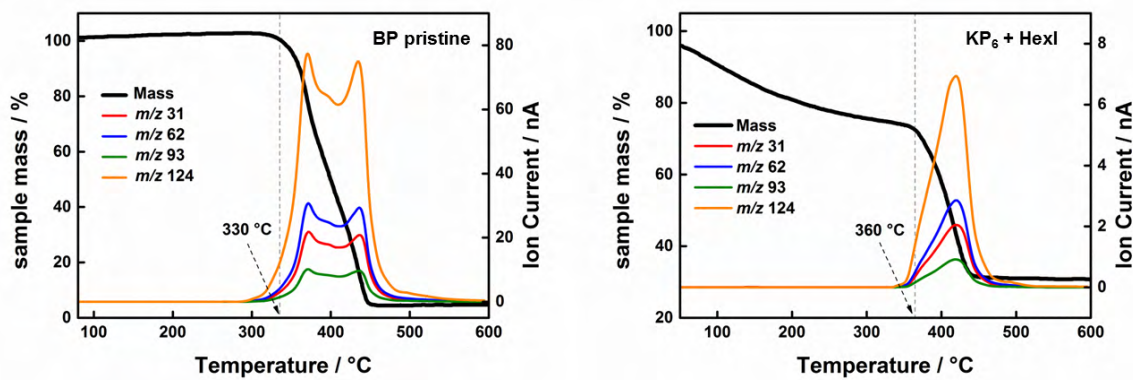

**Figure SI 14:**

TGA-MS analysis featuring complete decomposition of BP to P$_4$ ($m/z$ = 124): Indeed, clustering of single phosphorus atoms forming P$_2$ ($m/z$ = 62) and P$_3$ ($m/z$ = 93) can be observed. Interestingly, the onset of this sharp mass loss is shifted from 330 °C to 360 °C in the covalently modified sample (right) compared to pristine BP (left), indicating a better thermal stability.

**SI 15: Reductive Covalent Functionalization of KP$_6$ + hexyl iodide – TGA analysis (graphitization)**

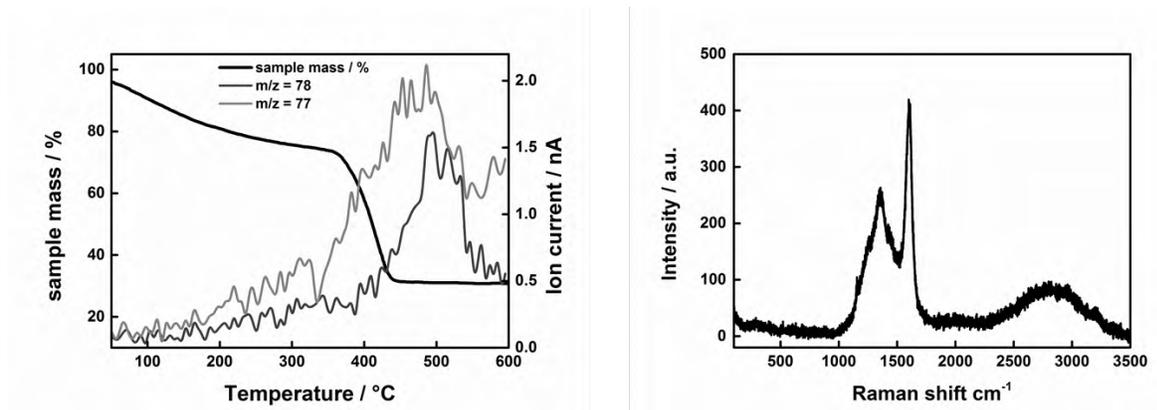

**Figure SI 15:**

(Left) TGA-MS analysis of KP$_6$ + hexyl iodide taking into account possible graphitization reactions above 400 °C: Indeed a significant amount of C$_6$H$_6$ as well C$_6$H$_5$ can be detected implying the graphitization of hexyl chains which were trapped before in between the layers. (Right) Mean Raman spectrum of the residual material obtained after the heating in the TGA up to 600 °C, clearly displaying graphitic features.

**SI 16: Temperature Dependent Raman Spectroscopy of KP$_6$ functionalized with hexyl iodide**

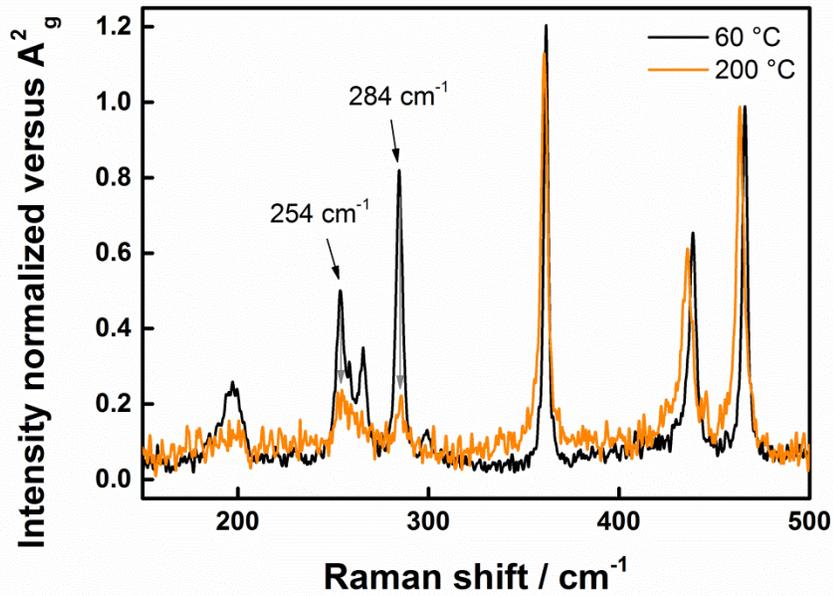

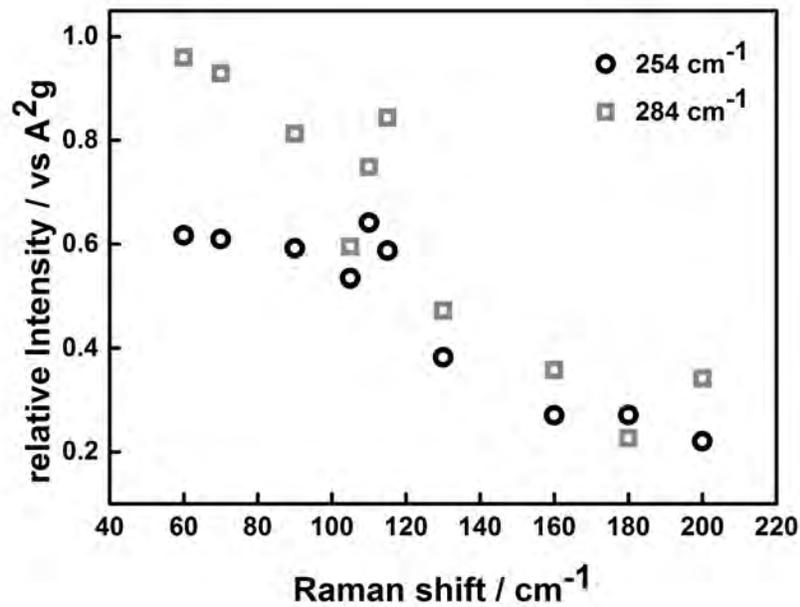

**Figure SI 16:**

Detailed analysis of the temperature-dependent Raman of KP$_6$ functionalized with hexyl iodide: (Top) Comparison of the Raman spectra recorded at 60 °C and 200 °C, normalized versus the $A^2_g$ vibrational mode of BP. (Bottom) By normalizing the recorded spectra versus the $A^2_g$ mode the relative decrease of the peaks at 284 cm$^{-1}$ and 254 cm$^{-1}$ –which can be related to the covalent functionalization of the BP lattice– can be seen more easily.

**SI 17: SRM of covalently functionalized BP *via* neutral route on Si/SiO$_2$**

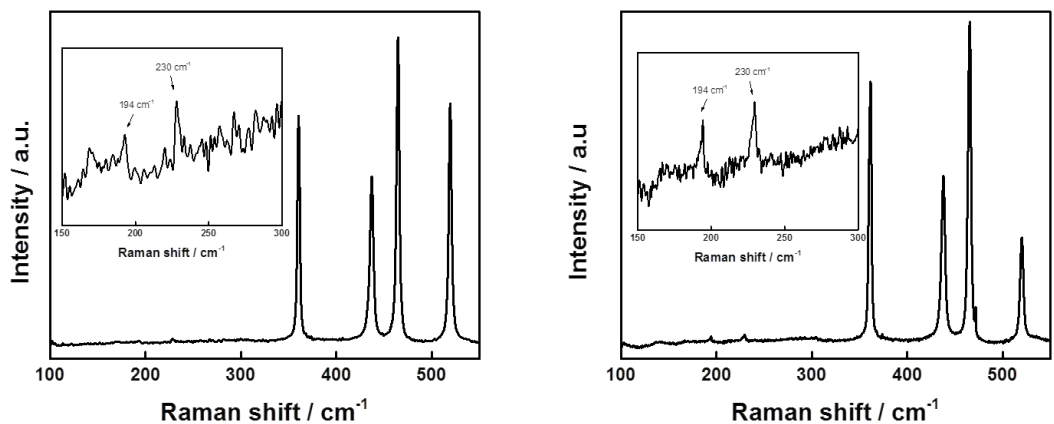

**Figure SI 17:**

Mean Raman spectra of BP + hexyl iodide in NMP reacted at room temperature (left) and at 100 °C (right). As it can be seen in both insets, only additional peaks at 194 cm$^{-1}$ and 230 cm$^{-1}$ (B$_{3g}$ and B$_{1g}$ modes) can be found.

**SI 18: TGA-MS analysis of covalently functionalized BP *via* neutral route**

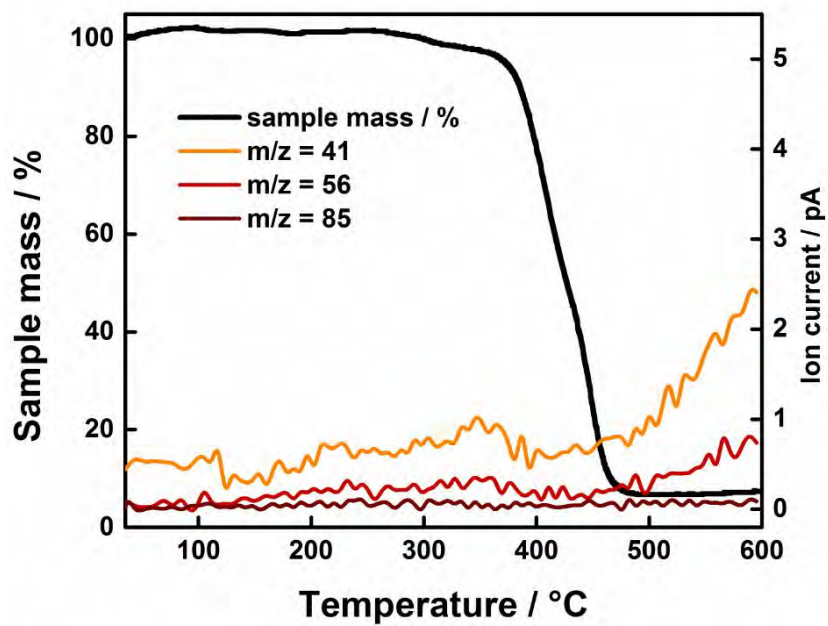

**Figure SI 18:**

TGA-MS analysis of BP + hexyl iodide reacted in NMP at room temperature: No evident mass loss between 100 °C and 300° can be seen, and also no characteristic fragments of the hexyl addend are detected in the MS indicating that the degree of functionalization of BP through the neutral route is very low.

**SI 19**: XPS survey spectra of pristine BP and covalently modified KP$_6$ with hexyl iodide

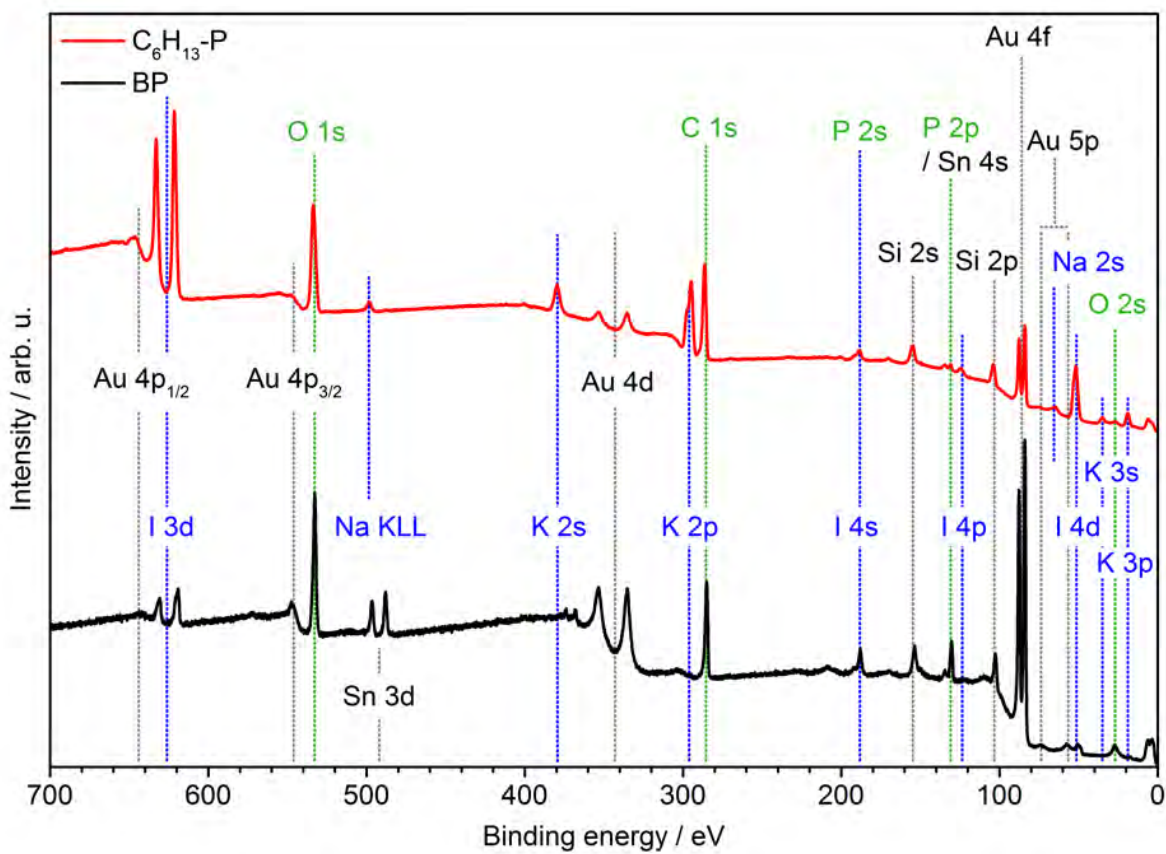

**Figure SI 19:**
XPS Survey spectra of the pristine BP (black), and C$_6$H$_{13}$-functionalized BP (red).

## SI 20: Raman Spectroscopy – Stability studies of $KP_6$ functionalized with hexyliodide

In order to test the stability of BP functionalized *via* reductive route SRS was conducted over time of a bulk sample thicker >100 nm. Along this front the absolute intensity of the $A^2_g$ vibrational mode was followed.

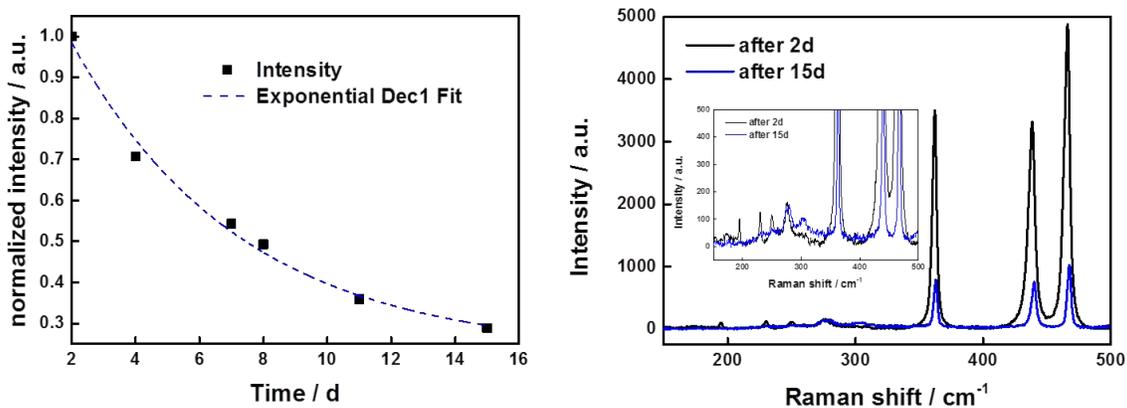

**Figure SI 20:**

(Left) Raman intensity of the $A^2_g$ vibrational Raman mode over time of $KP_6$ functionalized with hexyliodide. The intensity loss nicely follows an exponential decay typical for BP; however the sample is only stable up to 15 days. Thus, our results suggest that covalent functionalization has no positive on the passivation of BP. (Right) Comparison of mean Raman spectra of $KP_6$ functionalized with hexyliodide after 2 days and after 15 days. The inset highlights the presence of Raman modes occurring due to the covalent functionalization of the sample.

**SI 21: Water assisted de-functionalization of BP followed by XPS**

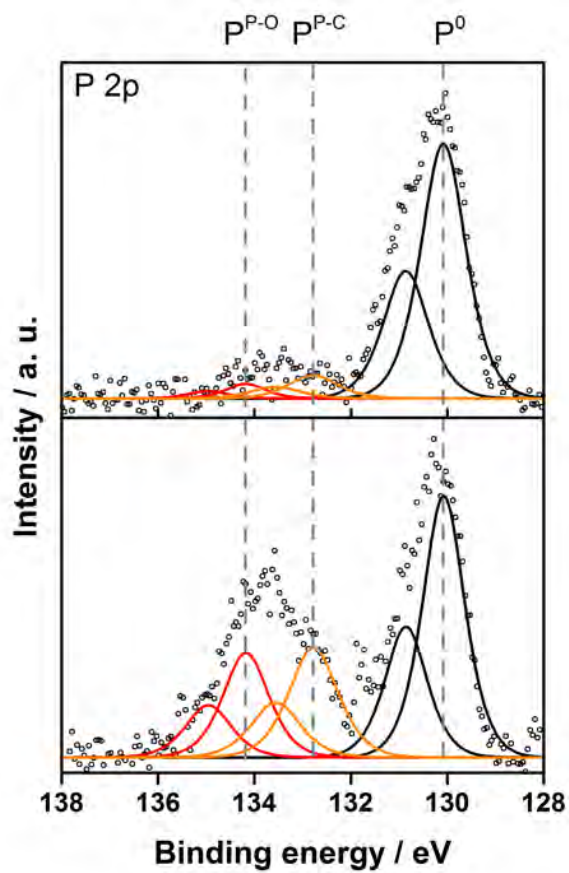

**Figure SI 21:**

XPS region scans of the P 2p region of hexyl-functionalized BP (bottom), and $H_2O$-rinsed hexyl-functionalized BP (top). Upon $H_2O$-exposure signals ascribed to oxygen-bound and functionalized phosphorus strongly decrease in intensity, indicating successful de-functionalization.

# SI 22: $^{31}$P-MAS NMR Spectroscopy

As a reference experiment, we measured quantitative $^{31}$P-MAS NMR of neutral BP with MeI in THF. Overall, the functionalization degree is determined to be less than 1% (no obvious shoulder at 22 ppm) upon de-convolution, which is significantly lower than for the intercalated sample.

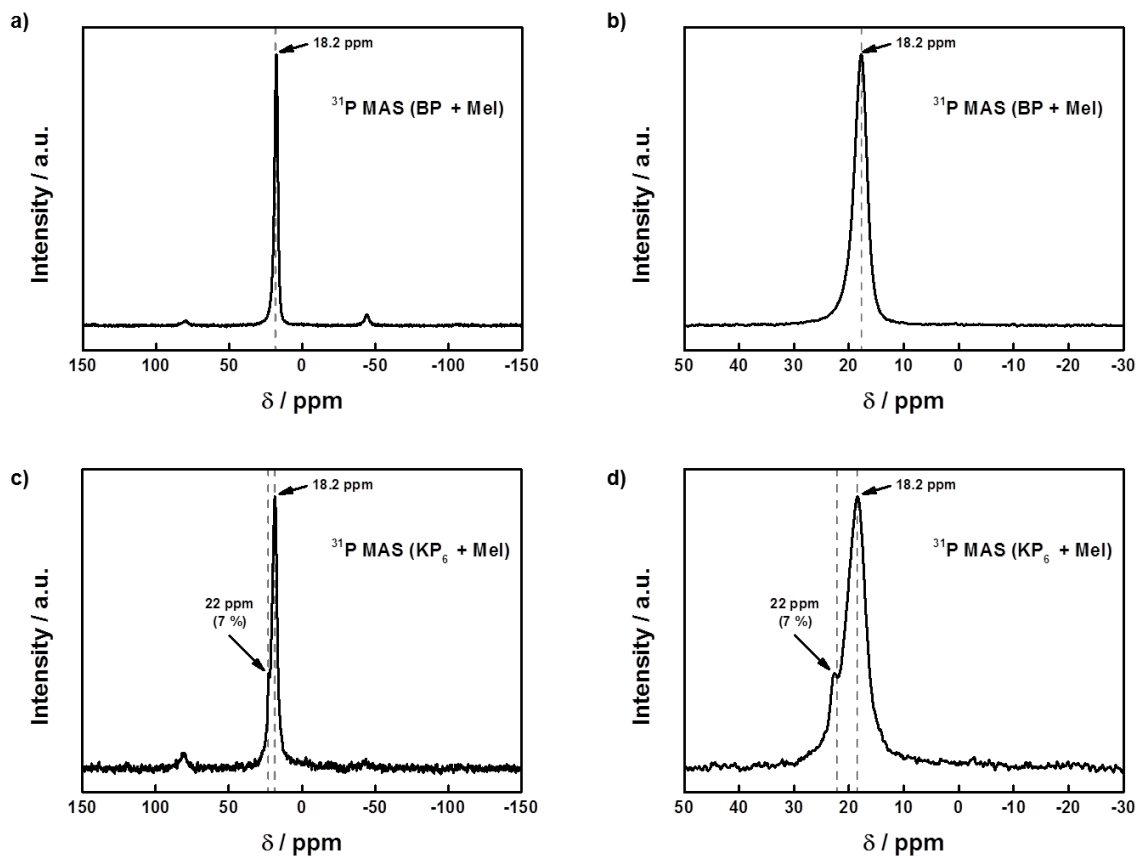

**Figure SI 22:**
(a) $^{31}$P-MAS NMR spectrum of neutral BP with MeI in THF displaying the complete range and all associated spinning side bands. (b) $^{31}$P-MAS NMR spectrum of neutral BP with MeI in THF showing a zoom of the area relevant for functionalization. (c) $^{31}$P-MAS NMR spectrum of KP$_6$ with MeI in THF displaying the complete range and all associated spinning side bands. Please note that the signal for intercalated KP$_6$ at -117 ppm is absent in the reaction product. (d) $^{31}$P-MAS NMR spectrum of neutral BP with MeI in THF showing a zoom of the area relevant for functionalization. Compared to the neutral route (BP with MeI in THF) a clear peak at 22 ppm can be observed, confirming the higher reactivity of the intercalated material KP$_6$ towards the alkyl halide than pristine BP.

**SI 23: UV-Vis Spectroscopy – Solubility studies**

We have evaluated the solubility of our samples in o-DCB after a short bath sonication for 10 min followed by centrifugation at 5000 rpm for 5 min by measuring UV-Vis. The concentration of the different samples were obtained after calculation of the extinction coefficient for dispersed BP ($\varepsilon_{367}$ = 1203 L g$^{-1}$ m$^{-1}$). The resulting solubility values (12–76 µg mL$^{-1}$) are in the same order of magnitude as reported for functionalized graphene (see Table S1).

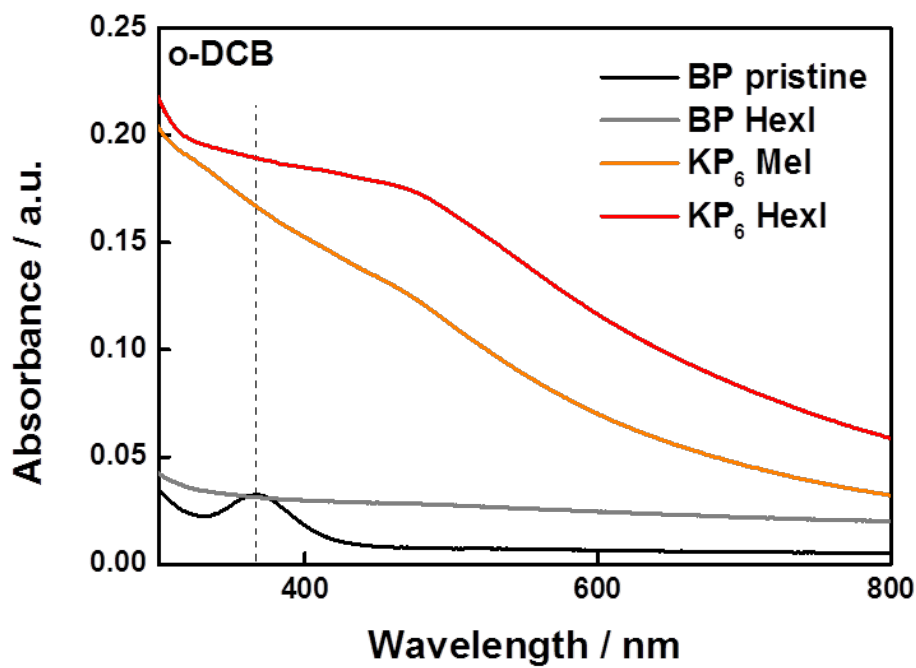

**Figure SI 23:**

UV-Vis spectra of supernatant solutions after centrifugation at 500 rpm for 5 min of pristine BP and its corresponding functionalized samples in o-DCB.

**Table S3:** Solubility values obtained for all samples in o-DCB in comparison to graphite

| Sample | Solubility / µg mL$^{-1}$ |
|---|---|
| pristine BP | 13.1 |
| BP + HexI (neutral) | 12.6 |
| KP$_6$ + MeI | 67.8 |
| KP$_6$ + HexI | 76.9 |
| pristine graphite | 16.0 |
| KC$_8$ + HexI | 72.0 |